\documentstyle[12pt]{article}

\newcommand {\nn}    {\nonumber}
\newcommand {\vs}[1]  { \vspace*{#1 cm} }

\newcounter{eq}
\newcounter{sc}

\newcommand {\MPL}  {Mod.Phys.Lett.}
\newcommand {\NP}   {Nucl.Phys.}
\newcommand {\PL}   {Phys.Lett.}
\newcommand {\PR}   {Phys.Rev.}
\newcommand {\PRL}   {Phys.Rev.Lett.}
\newcommand {\CMP}  {Comm.Math.Phys.}

\newcommand {\PTP}  {Prog.Theor.Phys.}

\def\overleftrightarrow#1{\vbox{\ialign{##\crcr
 $\leftrightarrow$\crcr\noalign{\kern-1pt\nointerlineskip}
 $\hfil\displaystyle{#1}\hfil$\crcr}}}

\newlength{\minitwocolumn}
\begin{document}


\begin{flushright}
EDO-EP-16\\
October, 1997\\
\end{flushright}
\vspace{30pt}

\pagestyle{empty}
\baselineskip15pt

\begin{center}
{\large\bf Supersymmetric IIB Matrix Models from \\
Space-time Uncertainty Principle and Topological Symmetry \vskip 1mm
}

\vspace{20mm}

Ichiro Oda
          \footnote{
          E-mail address:\ ioda@edogawa-u.ac.jp
                  }
\\
\vspace{10mm}
          Edogawa University,
          474 Komaki, Nagareyama City, Chiba 270-01, JAPAN \\

\end{center}


\vspace{15mm}
\begin{abstract}
Starting with topological field theory, we derive space-time uncertainty
relation proposed by Yoneya through breakdown of topological symmetry
in the large $N$ matrix model. 
Next, on the basis of only two basic principles, those are, generalized 
space-time uncertainty principle containing spinor field and topological 
symmetry, we construct a new matrix model. 
If we furthermore impose a requirement of $N=2$ supersymmetry, this new 
matrix model exactly reduces to the IKKT model or the Yoneya model for IIB
superstring depending on an appropriate choice for a scalar function.
A key feature of these formulations is an appearance of the nontrivial 
"dynamical" theory through breakdown of topological symmetry 
in the matrix model.
It is closely examined why the nontrivial "dynamical" theory appears
from the trivial topological field theory.

\vspace{15mm}

\end{abstract}

\newpage
\pagestyle{plain}
\pagenumbering{arabic}


\rm
\section{Introduction}

One of the most exciting achievements for theoretical physicists
is surely to construct a theory that explains all the experimental
data observed by then and predicts still unknown phenomena existing
in nature by starting with few fundamental principles deductively.
As a representative example of such theories, we are familiar with
general relativity by Einstein \cite{Ein}.
Even if general relativity is built from only two basic principles, 
namely, equivalence principle and general coordinate invariance 
by help of Riemannian geometry, it has explained not only all the
physical facts relevant to gravity but also predicted various 
remarkable things such as the gravitational redshift, the precession
of planetary orbits and the bending of light as well as an existence
of black holes \cite{Wald}.

It is nowadays widely expected that string theory \cite{GSW} might 
be the final theory unifying all the interactions among elementary
particles including the gravitational interaction. Then, it is a 
fascinating enterprise to try to construct string theory or M-theory 
\cite{MT} from few fundamental principles like general relativity.
However, at the present stage it is a very pity that our understanding
of the fundamental principles in string theory is far from complete.
Actually, in string theory we have a good grasp of neither the 
principle corresponding to the equivalence principle nor the 
gauge symmetry corresponding to the general covariance in 
comparison with general relativity.

Recently we have had some remarkable progress in non-perturbative 
formulations of M-theory \cite{M} and IIB superstring \cite{IKKT}.
These studies have provided us with an important clue to understand
the fundamental degrees of freedom at the short distance in a theory
containing gravity, where D-particles in M-theory and D-instantons
in IIB superstring constitute the fundamental building blocks for
membrane and string, respectively. However, from these studies
it seems to be difficult to get useful informations directly about 
the underlying fundamental principle and gauge symmetry behind 
M-theory and string theory.

On the other hand, in a quest of the fundamental principle of string
theory, Yoneya has advocated, what we call, the space-time uncertainty 
principle of string with respect to the time interval and the spacial
length, which has the form \cite{Y1,Y2}
\begin{eqnarray}
\Delta T \Delta X 
\ge l_s^2,
\label{1.1}
\end{eqnarray}
where $l_s$ denotes the string minimum length which is related
to the Regge slope $\alpha'$ by $l_s = \sqrt{\alpha'}$.
The space-time uncertainty principle (\ref{1.1})
would produce an interesting
physical picture that in string theory, maybe also in M-theory,
space-time in itself is quantized at the short distance and the
concept of space-time as a continuum manifold cannot be extrapolated
beyond the fundamental string scale $l_s$. It is also important 
to point out that this principle seems to be consistent with the
recent non-perturbative formulations of M-theory \cite{M} and IIB 
superstring \cite{IKKT} where this principle is realized 
implicitly in the form of the noncommutative geometry. 
Moreover, in terms of the "conformal constraint" coming from
the Schild action \cite{Schild} and essentially expressing
the space-time uncertainty priciple (\ref{1.1}), Yoneya \cite{Y2}
has constructed a IIB matrix model from which the IKKT model
\cite{IKKT} can be induced as an effective theory
for D-branes \cite{Pol}.

Being stimulated by Yoneya's works \cite{Y1,Y2}, in a preliminary 
study \cite{Oda} we have recently constructed a bosonic matrix model 
and shown that the equation of motion precisely describes a 
stronger form of the space-time uncertainty principle (\ref{1.1}). 
A key idea of this construction is to start with the topological 
field theory \cite{Witten1}, break this huge symmetry and then 
give rise to a nontrivial dynamical matrix theory whose moduli 
space is equal to a stronger form of the space-time uncertainty 
relation (\ref{1.1}). The aim of this article is not only to
present the full details of this preliminary study \cite{Oda} 
but also to generalize the results obtained there to a supersymmetric 
case in order to build matrix models for IIB superstring. 
As is well known at the moment, the supersymmetry is an essential 
ingredient in the recent 
development of the non-perturbative matrix models \cite{M,IKKT}
since the D-particle and the D-instanton are the BPS states preserving 
half of the supersymmetry and the supersymmetry guarantees the 
cluster property of these states. 

It should be emphasized that our goal in this paper is to explore
a possibility of formulating a non-perturbative string theory
from the first principles. As the first principles, we shall
take the space-time uncertainty principle and the topological
symmetry since the former principle describes a peculiar feature
of string theory and seems to be consistent with thought 
experiments done until now. On the other hand, although the
latter principle is still conjectural in string theory,
it is very appealing from the following arguments. 
A string has an infinite number of states in the perturbative
level in addition to various extended objects as solitonic 
excitations in the non-perturbative regime. Thus the 
local symmetry behind a string theory must be quite huge 
such that it controls so many states simultaneously 
without reference to their masslessness or massiveness.
The topological symmetry is a maximum local symmetry so
that it would be a strong candidate as such a huge local
symmetry.
 
Frankly speaking, however, at present we have no idea whether 
these two basic principles are really deep principles like 
the equivalence principle and the general covariance in 
general relativity or are just useful technical tools for 
construction of matrix models. Incidentally,
as for the topological symmetry, it would be worthwhile to point 
out that it has been already stated that the topological symmetry 
might be of critical importance in both string theory and quantum 
gravity in  connection with the background independent formulation 
of string theory and the unbroken phase of quantum gravity 
\cite{Witten2}.

The paper is organized as follows. In section 2 we briefly review 
Yoneya's works \cite{Y2} which are relevant to the present study.
Specific attention is paid to the "conformal" constraint and
the space-time uncertainty principle. In section 3, we derive 
a stronger form of the space-time uncertainty principle from the 
topological field theory where the classical action is trivially 
zero. The key idea here is the breakdown of the topological symmetry
in changing from the continuous field theory to the discrete
matrix model. In section 4, we incorporate the spinors in
the above theory and construct a new matrix model. If we require 
this theory to be invariant under $N=2$ supersymmetric 
transformations in ten dimensions, it turns out that this new 
matrix model becomes the IKKT model or the Yoneya model for type 
IIB superstring. This choice is dependent on the form of a 
classical solution for a scalar function.
The final section is devoted to discussions.

\section{ The conformal constraint and the space-time \\
           uncertainty principle }

In this section, we review only a part of Yoneya's works relevant
to later study (See \cite{Y1,Y2} for more details). Let us
start with the Schild action \cite{Schild} of a bosonic string.
Then the Schild action has the form
\begin{eqnarray}
S_{Schild} = -\frac{1}{2} \int d^2 \xi \left[\ -\frac{1}{2 \lambda^2}
\frac{1}{e} \left( \varepsilon^{ab} \partial_a X^\mu \partial_b
X^\nu \right)^2 + e \ \right],
\label{2.1}
\end{eqnarray}
where $X^\mu (\xi)$ $(\mu = 0, 1, \ldots , D-1)$ are space-time 
coordinates, $e(\xi)$ is a positive definite scalar density defined 
on the string world sheet parametrized by $\xi^1$ and $\xi^2$, 
and $\lambda = 4\pi \alpha'$. Throughout this paper, we assume that
the space-time metric takes the flat Minkowskian form defined as 
$\eta_{\mu\nu} = diag(- + + \ldots +)$.

Taking the variation
with respect to the auxiliary field $e(\xi)$, one obtains
\begin{eqnarray}
e(\xi) = \frac{1}{\lambda} \sqrt{-\frac{1}{2} 
\left( \varepsilon^{ab} \partial_a X^\mu \partial_b
X^\nu \right)^2 },
\label{2.2}
\end{eqnarray}
which is also rewritten to be
\begin{eqnarray}
\lambda^2 = -\frac{1}{2}  
\left\{  X^\mu, X^\nu \right\}^2,
\label{2.3}
\end{eqnarray}
where one has introduced the diffeomorphism invariant Poisson
bracket defined as 
\begin{eqnarray}  
\left\{  X^\mu, X^\nu \right\} = 
\frac{1}{e(\xi)} \varepsilon^{ab} \partial_a X^\mu \partial_b
X^\nu.
\label{2.4}
\end{eqnarray}
Then eliminating the auxiliary field $e(\xi)$ from (\ref{2.1}) through 
(\ref{2.2}) and using the identity
\begin{eqnarray}  
- \det \partial_a X \cdot \partial_b X
= -\frac{1}{2} \left( \varepsilon^{ab} \partial_a X^\mu \partial_b
X^\nu \right)^2,
\label{2.5}
\end{eqnarray}
the Schild action (\ref{2.1}) becomes at least classically 
equivalent to the famous Nambu-Goto action $S_{NG}$
\begin{eqnarray}
S_{Schild} &=& -\frac{1}{\lambda} \int d^2 \xi \sqrt{- \det 
\partial_a X \cdot \partial_b X},\nn\\
&=& S_{NG}.
\label{2.6}
\end{eqnarray}

In order to check that the "conformal" constraint (\ref{2.3}) expresses
half of the classical Virasoro conditions, it is convenient to use the 
Hamiltonian formalism \cite{Y2}. If we denote the differentiation
with respect to $\xi^1$ and $\xi^2$ by the dot and the prime,
respectively,
the canonical conjugate momenta to the $X^\mu$ are given by
\begin{eqnarray}
P^\mu = \frac{1}{\lambda^2} \frac{1}{e} \left( \dot{X^\mu} 
X'^2 - X'^\mu \dot{X} \cdot X' \right),
\label{2.7}
\end{eqnarray}
from which we can obtain the conventional classical Virasoro constraints
\begin{eqnarray}
P \cdot X' = 0, \\
P^2 + \frac{1}{\lambda^2} X'^2 = 0,
\label{2.8}
\end{eqnarray}
where the lapse constraint (10) is a consequence of the 
"conformal" constraint (\ref{2.3}) while the shift constraint
(9) comes from (8) trivially.

Let us clarify the physical implication of the "conformal"
constraint (\ref{2.3}). We are now familiar with the well-known relation 
between the Poisson bracket and the commutation relation in the 
large $N$ matrix model:
\begin{eqnarray}
\left\{ A, B \right\} \longleftrightarrow 
\left[ A, B \right].
\label{2.9}
\end{eqnarray}
Then the "conformal" constraint (\ref{2.3}) becomes 
\begin{eqnarray}
\lambda^2 = -\frac{1}{2}  
\left[  X^\mu, X^\nu \right]^2.
\label{2.10}
\end{eqnarray}
Recalling that the target space-time metric is now assumed to be
$\eta_{\mu\nu} = diag(- + + \ldots  +)$, Eq.(\ref{2.10}) can be
rewritten to be
\begin{eqnarray} 
\left[  X^0, X^i \right]^2 - \lambda^2 = \frac{1}{2}
\left[  X^i, X^j \right]^2,
\label{2.11}
\end{eqnarray}
where the summation over the transverse indices $i, j$ is taken.
The right-handed side of Eq.(\ref{2.11}) is a positive definite
hermitian operator, so this equation implies the space-time 
uncertainty principle (\ref{1.1}) under an appropriately defined
expectation value \cite{Y2}.
In this respect it is interesting to note that
the signature of the space-time must be not Euclidean but Minkowskian 
in order to get the space-time uncertainty principle (\ref{1.1}) from 
(\ref{2.10}). This point might give one justifiable reasoning to 
the problem of space-time signature \cite{Duff}.

As shown above, since a feature of the classical conformal invariance 
is contained in the space-time uncertainty principle, it is 
natural to postulate that non-perturbative string theory 
should be formulated on the basis of this principle \cite{Y2}.
Actually, Yoneya has derived such an action
which has a close connection with the IKKT model \cite{IKKT}.
His construction of the action is in itself quite interesting 
but a little ambiguous and ad hoc. In particular, it is unclear
what underlying symmetry exists behind the space-time uncertainty
principle.
In the next section, we shall take a different path of 
thought where we start with a topological field theory, from which 
we will derive the space-time uncertainty principle proposed by Yoneya
through the breakdown of the topological symmetry in the large $N$ 
matrix model. 
This derivation might suggest that the topological symmetry
is the underlying fundamental symmetry behind the space-time 
uncertainty principle of string theory.

\section{ A bosonic matrix model }

In this section let us construct a bosonic matrix model which 
expresses an essential content of the space-time uncertainty
principle. The preliminary report was given in the ref.\cite{Oda}.
Let us start by considering a topological theory \cite{Witten1}
where the classical action is trivially zero but dependent on
the fields $X^\mu(\xi)$ and $e(\xi)$ as follows:
\begin{eqnarray}
S_{c} = S_{c}(X^\mu(\xi), e(\xi)) = 0.
\label{3.1}
\end{eqnarray}
The BRST transformations corresponding to the topological
symmetry are given by
\begin{eqnarray}
\delta_B X^\mu = \alpha^\mu,  \ \delta_B \alpha^\mu = 0, \nn\\
\delta_B e = e \ \eta,  \ \delta_B \eta = 0, \nn\\
\delta_B \bar{c} = b,  \ \delta_B b = 0,
\label{3.2}
\end{eqnarray}
where $\psi^\mu$ and $\eta$ are ghosts, and $\bar{c}$ and $b$
are respectively an antighost and an auxiliary field. Note that
these BRST transformations are obviously nilpotent. Also notice
that the BRST transformation $\delta_B e$ shows the character as a
scalar density of $e$.

The idea, then, is to fix partially the topological symmetry
corresponding to $\delta_B e$ by introducing an appropriate
covariant gauge condition. A conventional covariant and 
nonsingular gauge condition would be $e = 1$ but this gauge
choice is not suitable for the present purpose since it 
makes difficult to pass to the large $N$ matrix theory. 
Then it is easy to check that
if we demand the space-time covariance almost the unique choice 
up to its polynomial forms is nothing but the "conformal" 
constraint (\ref{2.3}). Of course, there is an ambiguity whether the
fundamental parameter $\lambda$ must be nonzero or not from
the viewpoint of the IKKT matrix model. To the problem we 
have the following opinions. Firstly, nonzero $\lambda$ is 
more general than zero $\lambda$. Secondly, suppose that we
have fine-tuned $\lambda$ to be zero at the outset. But 
renormalization usually introduces such a dimensionful 
quantity into the quantum theory so that it is natural to 
include nonzero $\lambda$
in the gauge condition from the beginning.
Consequently the quantum action defined as $S_b = \int d^2 
\xi \ e L_b$ becomes
\begin{eqnarray}
L_b &=&  \frac{1}{e} \delta_B \left[ \bar{c} \left\{ e \left(
\frac{1}{2} \left\{ X^\mu, X^\nu \right\}^2 + \lambda^2 \right)
\right\} \right],\nn\\
&=&  b \left( \frac{1}{2} \left\{ X^\mu, 
X^\nu \right\}^2 + \lambda^2 \right)
- \bar{c} \left( \eta \left( -\frac{1}{2} \left\{ X^\mu, 
X^\nu \right\}^2 + \lambda^2 \right) 
+ 2 \left\{ X^\mu, X^\nu \right\} \left\{ X^\mu, \alpha^\nu 
\right\} \right),
\label{3.3}
\end{eqnarray}
where the BRST transformations (\ref{3.2}) were used. Here for later
convenience it is useful to redefine the auxiliary field $b$ by 
$b + \bar{c} \ \eta$. Then $L_b$ can be cast into a simpler form
\begin{eqnarray}
L_b =   b \left( \frac{1}{2} \left\{ X^\mu, 
X^\nu \right\}^2 + \lambda^2 \right)
- 2 \lambda^2 \bar{c} \ \eta - 2 \bar{c} \left\{ X^\mu, X^\nu \right\} 
\left\{ X^\mu, \alpha^\nu \right\}.
\label{3.4}
\end{eqnarray}

What is necessary to obtain a stronger form of the space-time
uncertainty relation (\ref{2.10}) is to change to the large $N$ matrix
theory where in addition to (\ref{2.9}) we have the following
correspondence
\begin{eqnarray}
\int d^2 \xi \ e \longleftrightarrow Trace,\nn\\
\int {\it D} e \longleftrightarrow \sum_{n=1}^\infty,
\label{3.5}
\end{eqnarray}
where the trace is taken over $SU(n)$ group. These
correspondence can be justified by expanding the
hermitian matices by $SU(n)$ generators in the large
$N$ limit as is reviewed by the reference \cite{F}.
We will discuss this point in detail in the final section.
Hence, for a moment, we assume that these correspondence is 
valid in our model.

Now in the large $N$ limit, we have
\begin{eqnarray}
S_b = Tr \left( b \left( \frac{1}{2} \left[ X^\mu, 
X^\nu \right]^2 + \lambda^2 \right)
- 2 \lambda^2 \bar{c} \ \eta - 2 \bar{c} \left[ X^\mu, X^\nu \right] 
\left[ X^\mu, \alpha^\nu \right] \right).
\label{3.6}
\end{eqnarray}
Then the partition function is defined as
\begin{eqnarray}
Z &=& \int {\it D}X^\mu {\it D}\alpha^\mu {\it D}e {\it D}\eta 
{\it D}\bar{c} {\it D}b \ e^{- S_b},\nn\\
&=& \sum_{n=1}^\infty \int {\it D}X^\mu {\it D}\alpha^\mu {\it D}\eta 
{\it D}\bar{c} {\it D}b \ e^{- S_b}.
\label{3.7}
\end{eqnarray}
At this stage, it is straightforward to perform the path integration
over $\eta$ and $\bar{c}$. Consequently, one obtains
\begin{eqnarray}
Z = \sum_{n=1}^\infty \int {\it D}X^\mu {\it D}\alpha^\mu {\it D}b \
e^{- Tr \ b \ \left( \frac{1}{2} \left[ X^\mu, X^\nu \right]^2 
+ \lambda^2 \right)}.
\label{3.8}
\end{eqnarray}
In (\ref{3.8}) since the quantum action does not depend on $\alpha^\mu$ 
it is obvious that there remains the gauge symmetry
\begin{eqnarray}
\delta \alpha^\mu = \omega^\mu,
\label{3.9}
\end{eqnarray}
which is of course nothing but the remaining topological symmetry. 
Now let us factor out this gauge volume or equivalently fix this 
gauge symmetry by the gauge condition $\alpha^\mu = 0$, so that 
the partition function is finally given by  
\begin{eqnarray}
Z = \sum_{n=1}^\infty \int {\it D}X^\mu {\it D}b \
e^{- Tr \ b \ \left( \frac{1}{2} \left[ X^\mu, X^\nu \right]^2 
+ \lambda^2 \right)}.
\label{3.10}
\end{eqnarray}

It is remarkable that the variation of the action with respect to the 
auxiliary variable $b$ in (\ref{3.10}) gives a stronger form of the 
space-time uncertainty relation (\ref{2.10}) and the theory is 
"dynamical" in the sense that
the ghosts have completely been decoupled in (\ref{3.10}). In other 
words, we
have shown how to derive the space-time uncertainty principle from
a topological theory through the breakdown of the topological symmetry
in the large $N$ matrix model. Why has the topological theory yielded
the nontrivial "dynamical" theory? The reason is very much simple.
In changing from the continuous theory (\ref{3.4}) to the matrix theory
(\ref{3.6}),
the dynamical degree of freedom associated with $e(\xi)$ was replaced
by the discrete sum over $n$ while the corresponding
BRST partner $\eta$ remains the continuous variable. This distinct
treatment of the BRST doublet leads to the breakdown of the topological
symmetry giving rise to a "dynamical" matrix theory. In this respect,
it is worthwhile to point out that while the topological symmetry
is "spontaneously" broken in this process, the other gauge symmetries 
never be violated (Of course, correctly speaking, 
these gauge symmetries reduce to the global symmetries in the
matrix model but this is irrelevant to the present argument).
Moreover, notice that the above-examined phenomenon is a peculiar
feature in the matrix model with the scalar density $e(\xi)$,
which means that an existence of the gravitational degree of
freedom is an essential ingredient since the generators of the
world-sheet reparametrizations, the Virasoro operators, provide
the Ward-identities associated with the target space general
covariance. 

\section{ Supersymmetric matrix models}

Having obtained a bosonic matrix model, we now turn our attention
to a more interesting model, i.e., its generalization to a 
supersymmetric matrix model. Actually, recent non-perturbative
formulations of M-theory \cite{M} and IIB superstring \cite{IKKT}
are based on the supersymmetry. Here we should
emphasize that our philosophy in constructing a supersymmetric
matrix model is rather different from the attitude in the bosonic
case in the previous section although we will go along a similar
path of procedure in what follows. Namely, so far by starting with
the topological field theory \cite{Witten1}, we have tried to derive
the space-time uncertainty principle proposed by Yoneya \cite{Y1,Y2}.
In this section, we promote the space-time uncertainty principle to
one of the basic principles for construction of a supersymmetric
matrix model. In other words, as mentioned in the abstract and the
introduction, on the basis of only two basic principles which are
the space-time uncertainty principle of string and the topological 
symmetry, we attempt to construct a new supersymmetric matrix model.
Of course, in the process of the model building, we will furthermore 
demand the invariance under the supersymmetric transformation.
Although the topological symmetry is broken (in some case even the 
space-time uncertainty principle is not explicit) at the final stage,
we will keep the strict invariance of a theory under the supersymmetry.
In this sense, at the present stage our basic principles might be 
interpreted as the starting principles for the model building.

As a first step for constructing a supersymmetric matrix model, one
has to require the classical action to depend on the Majorana spinor 
field $\psi_\alpha(\xi)$ as well as the bosonic fields $X^\mu(\xi)$ 
and $e(\xi)$
\begin{eqnarray}
S_{c} = S_{c}(X^\mu(\xi), \psi_\alpha(\xi), e(\xi)) = 0,
\label{4.1}
\end{eqnarray}
where the subscript $\alpha$ stands for spinor index which should
not be confused with the topological ghost $\alpha^\mu(\xi)$ 
corresponding to $X^\mu(\xi)$. The reason why we
consider only the Majorana spinor will be explained later. This
time, in addition to the BRST transformations (\ref{3.2}) one 
has to add the following BRST transformations for fermions:
\begin{eqnarray}
\delta_B \psi_\alpha = \beta_\alpha,  \ \delta_B \beta_\alpha = 0.
\label{4.2}
\end{eqnarray}

Next let us set up the gauge condition for $\delta_B e$. Instead
of the bosonic case
\begin{eqnarray}
\frac{1}{2} \left\{ X^\mu, X^\nu \right\}^2 + \lambda^2 = 0,
\label{4.3}
\end{eqnarray}
we shall set up its natural extension involving the spinor field
\begin{eqnarray}
\frac{1}{2} \left\{ X^\mu, X^\nu \right\}^2 + \lambda^2 
+ \frac{1}{2} \bar{\psi} \Gamma_\mu \left\{ X^\mu, \psi \right\}
= 0.
\label{4.4}
\end{eqnarray}
When transforming to the matrix theory later, this gauge condition
becomes a generalized stronger form of the space-time uncertainty 
principle. Although this generalized form is different from the
original one proposed by Yoneya \cite{Y1,Y2} by the spinor part,
in the ground state they are equivalent so we take the above
gauge condition (\ref{4.4}). Interestingly enough, it will be
shown later that the gauge choice (\ref{4.4}) leads to the same 
theory as Yoneya's one if a suitable solution for the auxiliary 
variable is chosen. Incidentally, the spinor part in
(\ref{4.4}) is adopted from an analogy with the supersymmetric 
Yang-Mills theory.
Thus we have the quantum action $S_q = \int d^2 \xi \ e \left( L_b 
+ L_f \right)$ with the bosonic contribution $L_b$ (\ref{3.3}) and the
fermionic one $L_f$ given by
\begin{eqnarray}
L_f &=&  \frac{1}{e} \delta_B \left( \bar{c} \ e \
\frac{1}{2} \bar{\psi} \Gamma_\mu \left\{ X^\mu, \psi \right\} 
\right),  \nn\\
&=&  b \ \frac{1}{2} \bar{\psi} \Gamma_\mu \left\{ X^\mu, \psi \right\}
- \bar{c} \ \frac{1}{2} \left( \bar{\beta} \Gamma_\mu \left\{ X^\mu,
\psi \right\} - \bar{\psi} \Gamma_\mu \left\{ \alpha^\mu, 
\psi \right\} - \bar{\psi} \Gamma_\mu \left\{ X^\mu, \beta \right\}
\right).
\label{4.5}
\end{eqnarray}
Here in a similar way to the bosonic case, let us redefine the
auxiliary field $b$ and the ghost $\beta$ by $b + \bar{c} \ \eta$
and $\beta - \frac{1}{2} \psi \ \eta$, respectively. As a result,
$L_b$ is given by (\ref{3.4}), on the other hand, $L_f$ takes the same 
form as (\ref{4.5}). When we rewrite the fermionic part $L_f$ in this
process, we need the famous Majorana identity $\bar{\psi} 
\Gamma_\mu \psi = 0$, for which we have confined ourselves to 
the Majorana spinor in this paper.  

As before, at this stage let us pass to the matrix model. Again
it is straightforward to carry out the path integration over
$\bar{c}$ and $\eta$ in a perfect analogous way to the bosonic
theory. Accordingly, we arrive at the following partition 
function
\begin{eqnarray}
Z = \sum_{n=1}^\infty \int {\it D}X^\mu {\it D}\alpha^\mu 
{\it D}\psi_\alpha {\it D}\beta_\alpha {\it D}b \
e^{- Tr \ b \ \left( \frac{1}{2} \left[ X^\mu, X^\nu \right]^2 
+ \lambda^2 + \frac{1}{2} \bar{\psi} \Gamma_\mu \left[ X^\mu, \psi 
\right] \right)}.
\label{4.6}
\end{eqnarray}
In this expression since the quantum action is independent of 
$\alpha^\mu$ and $\beta_\alpha$ we have the remaining topological
symmetries given by
\begin{eqnarray}
\delta \alpha^\mu = \omega^\mu, \ \delta \beta_\alpha = 
\rho_\alpha.
\label{4.7}
\end{eqnarray}
After factoring these gauge volumes out, the partition function is 
finally cast to be 
\begin{eqnarray}
Z &=& \sum_{n=1}^\infty \int {\it D}X^\mu {\it D}\psi_\alpha 
{\it D}b \ e^{- S_q}, \nn\\
&=& \sum_{n=1}^\infty \int {\it D}X^\mu {\it D}\psi_\alpha 
{\it D}b \
e^{- Tr \ b \ \left( \frac{1}{2} \left[ X^\mu, X^\nu \right]^2 
+ \lambda^2 + \frac{1}{2} \bar{\psi} \Gamma_\mu \left[ X^\mu, \psi 
\right] \right)}.
\label{4.8}
\end{eqnarray}
Of course, the action $S_q$ still possesses the zero volume
reduction of the usual gauge symmetry
\begin{eqnarray}
\delta \psi_\alpha &=& i \left[X_\mu, \Lambda \right], \nn\\
\delta X_\mu &=& i \left[\psi, \Lambda \right], \nn\\ 
\delta b &=& i \left[b, \Lambda \right].
\label{4.9}
\end{eqnarray}
And it is straightforward to derive the equations of
motion from $S_q$ whose results are written as
\begin{eqnarray}
\frac{1}{2} \left[X^\mu, X^\nu \right]^2 + \lambda^2
+ \frac{1}{2} \bar{\psi} \Gamma_\mu \left[X^\mu, 
\psi \right] &=& 0, \\
\left[X^\mu, b \left[ X_\mu, X_\nu \right] \right]
+ \frac{1}{4}  \left[ \ b \ \bar{\psi} \Gamma^\nu, 
\psi \right]_+ &=& 0, \\
\left[X^\mu, \Gamma_\mu \psi \right] b + \frac{1}{2} \Gamma_\mu \psi 
\left[X^\mu, b \ \right] &=& 0,
\label{4.10}
\end{eqnarray}
where $[ \ , \ ]_+$ denotes the anticommutator.

In this way, we have constructed a new matrix model with the 
Majorana spinor variable on the basis of the space-time uncertainty
principle and the topological symmetry. Although the action
contains the spinor variable in addition to the bosonic
variable, it is not always supersymmetric. The supersymmetry plays 
the most critical role in the matrix models for M-theory
\cite{M} and IIB superstring theory \cite{IKKT}, 
so we should require the invariance under the supersymmetry 
for the action $S_q$ obtained in (\ref{4.8}). The most natural 
form of $N=2$ supersymmetric transformations is motivated
by a supersymmetric Yang-Mills theory whose
(0+0)-dimensional reduction is given by
\begin{eqnarray}
\delta \psi_\alpha^{ab} &=& i \left[X_\mu, X_\nu \right]^{ab}
\left(\Gamma^{\mu\nu} \varepsilon \right)_\alpha + \zeta_\alpha
\delta^{ab}, \nn\\
\delta X_\mu^{ab} &=& i \bar{\varepsilon} \Gamma_\mu 
\psi^{ab}, \nn\\
\delta b^{ab} &=& 0,
\label{4.11}
\end{eqnarray}
where we have explicitly written down the matrix indices to
clarify that $\varepsilon_\alpha$ and $\zeta_\alpha$ are the 
Majorana spinor parameters. These supersymmetric transformations 
are of the same form as in IKKT model \cite{IKKT}. At this stage,
we assume the space-time dimensions to be ten in order to make contact 
with IIB superstring. 

To make the action $S_q$ in (\ref{4.8}) invariant under the $N=2$ 
supersymmetry (\ref{4.11}), it is easy to check that $b^{ab}$ must 
take the diagonal form with respect to the hermitian matrix indices. 
There are two interesting solutions. One of them is to select the 
auxiliary variable $b^{ab}$ to be proportional to $\delta^{ab}$ up
to a constant. Without generality we take the proportional constant
to be $- \frac{1}{2}$, therefore
\begin{eqnarray}
b^{ab} = - \frac{1}{2} \delta^{ab}.
\label{4.12}
\end{eqnarray}
Here if we redefine $X^\mu$, $\psi$, and $- \frac{1}{2} \lambda^2$
in terms of $\alpha^{\frac{1}{4}} X^\mu$, $\sqrt{2} 
\alpha^{\frac{3}{8}} \psi$, and $\beta$, respectively, the action 
$S_q$ can be rewritten to be
\begin{eqnarray}
S_q = \alpha \left(- \frac{1}{4} Tr \left[ X^\mu, X^\nu \right]^2 
- \frac{1}{2} Tr \bar{\psi} \Gamma_\mu \left[ X^\mu, \psi 
\right] \right) + \beta Tr \mathbf{1}.
\label{4.13}
\end{eqnarray}
Note that this action is completely equivalent to the action in
the IKKT model \cite{IKKT}. In this case, we cannot derive the
space-time uncertainty relation from the equation of motion,
but this relation might be encoded implicitly in the matrix 
character of the model.

The other interesting solution would be of the form
\begin{eqnarray}
b^{ab} = c \ \delta^{ab},
\label{4.14}
\end{eqnarray}
with some additional auxiliary variable $c$. With this choice, the
partition function (\ref{4.8}) can be reduced to be 
\begin{eqnarray}
Z &=& \sum_{n=1}^\infty \int {\it D}X^\mu {\it D}\psi_\alpha 
{\it D}c \ e^{- S_q}, \nn\\
&=& \sum_{n=1}^\infty \int {\it D}X^\mu {\it D}\psi_\alpha 
{\it D}c \
e^{- \ c \ Tr \left( \frac{1}{2} \left[ X^\mu, X^\nu \right]^2 
+ \lambda^2 + \frac{1}{2} \bar{\psi} \Gamma_\mu \left[ X^\mu, \psi 
\right] \right)}.
\label{4.15}
\end{eqnarray}
At first sight, it seems that we have obtained a new supersymmetric
matrix model, but this is an illusion. We shall show that the above
model is entirely equivalent to the Yoneya model \cite{Y2} in what 
follows. Provided that we take account of the stronger form of the 
space-time uncertainty principle instead of the weaker form, the 
Yoneya model can be expressed in terms of the partition function
\begin{eqnarray}
Z &=& \sum_{n=1}^\infty \int {\it D}X^\mu {\it D}\psi_\alpha 
{\it D}c \ e^{- S_y}, \nn\\
&=& \sum_{n=1}^\infty \int {\it D}X^\mu {\it D}\psi_\alpha 
{\it D}c \
e^{- \ c \ Tr \left( \frac{1}{2} \left[ X^\mu, X^\nu \right]^2 
+ \lambda^2 \right) - Tr \frac{1}{2} \bar{\psi} \Gamma_\mu 
\left[ X^\mu, \psi \right] }.
\label{4.16}
\end{eqnarray}
This partition in the Yoneya model does not look like the partition
(\ref{4.15}). But Yoneya has defined the
supersymmetric transformations in a slightly different manner
compared to ours (\ref{4.11}). His supersymmetry is
\begin{eqnarray}
\delta \psi_\alpha^{ab} &=& i c  \left[X_\mu, X_\nu \right]^{ab}
\left(\Gamma^{\mu\nu} \varepsilon \right)_\alpha + \zeta_\alpha
\delta^{ab}, \nn\\
\delta X_\mu^{ab} &=& i \bar{\varepsilon} \Gamma_\mu 
\psi^{ab}, \nn\\
\delta c &=& 0.
\label{4.17}
\end{eqnarray}
Note that there exists $c$ variable in the first term
of the right-handed side in the first equation while it is 
absent in our formula (\ref{4.11}) (Of course, in (\ref{4.11}) 
we should replace $\delta b^{ab} = 0$ with $\delta c = 0$ 
for present consideration). Then it is easy to show that if
we redefine $\psi$, $\varepsilon$ and $\zeta$ by
$c^{\frac{1}{2}} \psi$, $c^{-\frac{1}{2}} \varepsilon$ and 
$c^{\frac{1}{2}} \zeta$, respectively in the Yoneya model,
Yoneya's action $S_y$ and supersymmetric transformations
(\ref{4.17}) conform to our action $S_q$ and supersymmetric
transformations (\ref{4.11}), respectively. To demonstrate 
a complete equivalence, we have to consider the functional 
measure. From these redefinitions
the functional measure receives a contribution of an additional 
factor $c^8$, but this change is absorbed into a definition 
of the functional measure ${\it D}c$ since the variable $c$
is the supersymmetrically invariant non-dynamical auxiliary 
variable in the model at hand. 
In this way, we can show that the solution (\ref{4.14})
gives rise to the Yoneya model. It is surprising that depending
on a choice of the scalar function $b$ our model leads to
the IKKT model \cite{IKKT} and the Yoneya model \cite{Y2},
which on reflection clarifies the difference between both 
the matrix models.

\section{ Discussions }

In this article we have investigated mainly two problems. One of
them is a possibility of the space-time uncertainty principle 
advocated by Yoneya \cite{Y1,Y2} to be derived from the topological 
field theory \cite{Oda}.
This study suggests that the underlying symmetry behind
this principle in string theory might be a topological symmetry as
mentioned before in a different context \cite{Witten2}. 
The other problem is to derive the supersymmetric matrix models
from the first principles based on the space-time uncertainty
principle and the topological symmetry and examine the relation
between the matrix model obtained in this way and the known
matrix models. We have observed that our matrix model contains
both the IKKT model and the Yoneya model if we demand the 
supersymmetry.

A rather unexpected appearance of the topological field theory 
seems to be plausible from the following intuitive arguments. 
Suppose that we live in the world where the topological symmetry 
is exactly valid. 
In such a world we have no means of measuring any distance 
owing to lack of the metric tensor field so that there is neither 
concept of distance nor the space-time uncertainty principle. 
But once the topological symmetry which is particularly connected 
with the gravitational degrees of freedom, is spontaneously broken 
by some dynamical mechanism, 
an existence of the dynamical metric together with a string having 
the minimum length would give us both concepts of the distance and 
the space-time uncertainty principle. Our bosonic matrix model 
seems to realize this scenario in a concrete way. 

Here we would like to comment on one important problem. In our
models, as in the IKKT model \cite{IKKT} the matrix size $n$ is
now regarded as a dynamical variable so that the
partition function includes the summation over $n$.
Even if the direct proof is missing, the summation
over $n$ is expected to recover the path integration
over $e(\xi)$. In fact, the authors of the reference
\cite{FKKT} have recently shown that the model of
Fayyazuddin et al. \cite{F, Solo} where a positive definite
hermitian matrix $Y$ is introduced as a dynamical
variable instead of $n$, belongs to the same 
universality class as the IKKT model \cite{IKKT}
owing to irrelevant deformations of the loop equation
\cite{Kitazawa}. Thus we think that the correspondence
(\ref{2.10}) and (\ref{3.5}) are legitimate even in 
the context at hand. Related to this problem, there
is an interesting recent conjecture in the 
non-perturbative formulation of M-theory that the 
equivalence between M-theory and Matrix theory is not limited
to the large $N$ limit but is valid for finite $N$
\cite{Suss}.  More recently this conjecture has been
proved to be correct up to two loops by evaluating 
the effective action for the scattering of two
D0-branes \cite{Becker, Becker2}.

In the bosonic model, we have not paid attention to the number 
of the space-time dimensions. In fact any dimensions except  
$D < 2$ are allowed. But an intriguing case happens when $D = 2$ 
even if this specification is not always necessary within the 
formulation. In this special dimension, the Nambu-Goto action 
which is at least classically equivalent to the Schild action 
as shown in (\ref{2.6}) becomes not only the topological field 
theory but also almost a surface term as follows:
\begin{eqnarray}
\sqrt{- \det \partial_a X \cdot \partial_b X} 
&=& \sqrt{- \left( \det \partial_a X^\mu \right)^2}, \nn\\
&=& \pm \det \partial_a X^\mu, \nn\\
&=& \mp \frac{1}{2} \varepsilon^{ab} \varepsilon_{\mu\nu}
\partial_a X^\mu \partial_b X^\nu,
\label{5.1}
\end{eqnarray}
where we have assumed a smooth parametrization of $X^\mu$ over
$\xi^a$ in order to take out the absolute value. Actually, this
topological model has been investigated to some extent in the 
past \cite{Fuji, Roberto, Akama}. In this case it is interesting
that we can start with not zero but the nontrivial surface term
as a classical action.

Our approach heavily relies on the mechanism of the breakdown
of the topological symmetry, so we should examine more closely
the reason why our model gives rise to the nontrivial 
"dynamical" theory from at least classically trivial
topological theory.  As mentioned in section 3, the technical
reason lies in asymmetric treatment between the BRST doublet
$e$ and $\eta$. However, there exists a deeper reason behind
it. To make our arguments clear, it is useful to compare the 
present approach with the previous studies about the topological
(pregauge-) pregeometric models \cite{Roberto, Akama} whose 
essential ideas will be recapitulated in what follows.

For generality, we consider an arbitrary dimension of space-time. 
We take the Nambu-Goto action as a classical action 
where we restrict ourselves to the case that the dimension 
is equal between the world-volume and the space-time. 
Then in a similar argument to (\ref{5.1}) we can prove that 
this classical action becomes topological. This is because we
can eliminate all the dynamical degrees of freedom by means of
the world-volume reparametrizations.  Let us rewrite it
to the Polyakov form
\begin{eqnarray}
S &=& -\frac{1}{\lambda} \int d^D \xi \sqrt{- \det 
\partial_a X \cdot \partial_b X},\nn\\
&=& \int d^D \xi \sqrt{-g} \left( g^{ab} \partial_a X^\mu
\partial_b X^\nu + \lambda \right).
\label{5.2}
\end{eqnarray}
In spite of lack of proof, the above two actions might be 
equivalent even in the quantum level as well as the classical
level owing to the topological character where there is no
anomaly. Next work is to evaluate the effective action for
the metric $g^{ab}$ due to the quantum fluctuation of the
"matter" fields $X^\mu$ whose result is given by \cite{Tera}
\begin{eqnarray}
S_{eff} = i \ Tr \log \left[\left( \partial_a \sqrt{-g} g^{ab}
\partial_b \right)\right] +  \lambda \int d^D \xi \sqrt{-g}.
\label{5.3}
\end{eqnarray}
When the curvature is small, it reduces to the Einstein-Hilbert
action with the cosmological constant
\begin{eqnarray}
S_{eff} = \int d^D \xi \sqrt{-g} \left( \tilde{\lambda} 
+ \frac{1}{16 \pi G} R + O(R^2, \log \Lambda^2) \right),
\label{5.4}
\end{eqnarray}
with 
\begin{eqnarray}
\tilde{\lambda} &=& \frac{D \Lambda^4}{8 (4 \pi)^2} +
\lambda, \nn\\
\frac{1}{16 \pi G} &=& \frac{D \Lambda^2}{24 (4 \pi)^2},
\label{5.5}
\end{eqnarray}
where we have introduced the momentum cutoff $\Lambda$ of the
Pauli-Villars type. Note that (\ref{5.5}) shows that we can choose
the effective cosmological constant $\tilde{\lambda}$ as small 
as we want, and the cutoff $\Lambda$ is of the order the Planck 
mass. It is quite interesting to ask why the topological action
has produced the Einstein-Hilbert action. This is because
the momentum cutoff $\Lambda$ breaks the topological symmetry
with keeping the general covariance. In other words, we have 
secretly introduced seed for breaking the topological symmetry
by the form of the cutoff. Of course, it is an interesting
idea to make a conjecture that renormalization induces such a
scale, but it seems to be quite difficult to prove this conjecture.

{}From this point of view, it is valuable to reconsider why 
the present formulation has produced the nontrivial matrix models
from the topological field theory. Originally, in membrane 
world, the matrix model has appeared to regularize the lightcone
supermembrane action with area-preserving diffeomorphisms where it 
has been remarkably shown that the action becomes exactly that 
of ten dimensional $SU(n)$ supersymmetric Yang-Mills theory 
reduced to $(0+1)$-dimensions \cite{DeWitt}. Similarly,
in our models, changing from the continuous topological
field theory to the discrete matrix model is equal to an
introduction of the regularization where the regularization
parameter corresponds to the size of the matrices. 
This type of the regularization breaks
only the topological symmetry, from which we can obtain the
nontrivial "dynamical" matrix models. It is very interesting
that the matrix model is equipped with such a natural
regularization scheme in itself. If the topological symmetry
is truly broken by some mechanism in order to make the 
topological field theory a physically vital theory, we believe
that theories equipped with some natural regularization scheme 
such as matrix model and induced gravity (pregeometry) would  
play an important role. In connection with string theory with
$\frac{1}{N}$ expansion, it is remarkable that several years
ago Thorn has already made a
conjecture that the local theory underlying string theory
should be either a theory with no curvature terms, as in
induced gravity or a topological field theory \cite{Thorn}.
The present formalism realizes this conjecture to a certain
extent.

\vs 1
\begin{flushleft}
{\bf Acknowledgement}
\end{flushleft}
The author thanks Y.Kitazawa and A.Sugamoto for valuable discussions. 
He is also indebted to M.Tonin for stimulating discussions,
a useful information on ref.\cite{Solo} and a kind hospitality at 
Padova University where the former part of this work has been done. 
After putting
the short article \cite{Oda} on hep-th database, he has received 
several e-mails pointing out a relation between it and other works.
He would like to acknowledge their informations, in particular, 
M.Kato \cite{Kato}, C.Castro \cite{Castro}, M.Li \cite{Li}, and 
G.Amelino-Camelia \cite{Amelino}.
This work was supported in part by Grant-Aid 
for Scientific Research from Ministry of Education, Science and
Culture No.09740212.

\vs 1

\end{document}